\documentclass[conference,compsoc]{IEEEtran}
\ifCLASSOPTIONcompsoc
  \usepackage[nocompress]{cite}
\else
  \usepackage{cite}
\fi
\ifCLASSINFOpdf
\else
\fi
\usepackage{graphicx}
\usepackage{subfig}
\usepackage{url}
\usepackage{booktabs}
\usepackage{hyperref}
\usepackage{multicol}
\usepackage{caption}
\usepackage[T1]{fontenc}

\usepackage{amsmath}
\usepackage{fancyhdr} 
\usepackage{tikz}     

\usepackage{fixltx2e}
\fancypagestyle{firstpage}{ 
    \fancyhf{} 
    \fancyfoot[L]{\begin{tikzpicture}[remember picture,overlay]
        \node[anchor=south west, xshift=1.75cm, yshift=1cm] at (current page.south west) {\footnotesize 979-8-3315-1127-2/24/\$31.00 ©2024 IEEE};
    \end{tikzpicture}} 
}

\begin{document}

%
\title{Segmentation of Coronary Artery Stenosis in X-ray Angiography \\using Mamba Models
}

\author{\IEEEauthorblockN{Ali Rostami}
\IEEEauthorblockA{Department of Mathematics, Statistics\\ and Computer Science \\
College of Science,\\
University of Tehran\\
Tehran, Iran\\
ali.rostami.1999@ut.ac.ir}
\and
\IEEEauthorblockN{Fatemeh Fouladi}
\IEEEauthorblockA{Department of Mathematics, Statistics\\ and Computer Science \\
College of Science,\\
University of Tehran\\
Tehran, Iran\\
fatemefouladi@ut.ac.ir}
\and
\IEEEauthorblockN{Hedieh Sajedi}
\IEEEauthorblockA{Department of Mathematics, Statistics\\ and Computer Science \\
College of Science,\\
University of Tehran\\
Tehran, Iran\\
hhsajedi@ut.ac.ir}
}


%


\maketitle

\begin{abstract}
Coronary artery disease stands as one of the primary contributors to global mortality rates. The automated identification of coronary artery stenosis from X-ray images plays a critical role in the diagnostic process for coronary heart disease. This task is challenging due to the complex structure of coronary arteries, intrinsic noise in X-ray images, and the fact that stenotic coronary arteries appear narrow and blurred in X-ray angiographies. This study employs five different variants of the Mamba-based model and one variant of the Swin Transformer-based model, primarily based on the U-Net architecture, for the localization of stenosis in Coronary artery disease. Our best results showed an F1 score of 68.79\% for the U-Mamba BOT model, representing an 11.8\% improvement over the semi-supervised approach.

\end{abstract}


%
\IEEEpeerreviewmaketitle

\section{Introduction}
Coronary artery disease (CAD) is the foremost cause of mortality globally, responsible for claiming over 17 million lives annually \cite{abbas2022automatic}. CAD is essentially caused by the deposition of atherosclerotic plaque in epicardial arteries, which leads to a mismatch between the oxygen supply and demand in the myocardium, and usually increases the risk of ischemia \cite{libby2005pathophysiology}. This is usually when the coronary arteries, being some of the major blood vessels supplying blood and oxygen to the heart muscle, have become narrowed by the deposition and accumulation of fats, cholesterol, and other substances on the inner walls. This is also referred to as stenosis. This limits the amount of oxygen supplied to the heart to its oxygen demand, leading to ischemia. Heart failure and even death are the potential risks to the patient. CAD symptoms may include chest pain known as angina, shortness of breath and fatigue, which occur when the heart is deprived of an adequate supply of oxygen-rich blood.

Early diagnosis of stenosis is crucial, and timely identification will improve cardiac function with a reduced risk of future cardiovascular complications. The diagnosis of coronary artery stenosis before depended on the combination of both noninvasive and invasive modalities. Noninvasive imaging modalities, such as computed tomography coronary angiography and magnetic resonance imaging, have been able to yield details on the anatomy of the coronary arteries, hence detecting stenosis without catheterization. Until now, the assessment methods for measuring the effect of stenosis on myocardial blood flow and cardiac function have primarily been stress tests and nuclear medicine myocardial perfusion imaging.

In this paper, we use X-ray coronary angiography (XCA) images for detecting stenosis. To this day, coronary artery invasive angiography remains the gold standard for diagnosing stenosis in coronary arteries. This represents an invasive procedure in which a catheter is inserted into the coronary arteries and a dye is injected; the latter provides the ability to visualize the coronary arteries under X-ray imaging. This has allowed clinicians to identify and quantify with a high degree of accuracy the extent of arterial narrowing. Detailed visualization obtained by angiography makes the method very important in evaluating the extent of CAD and appropriately planning any intervention. On the other hand, coronary arteries having stenosis appear to be narrow and blurred in XCA so that they become hardly noticeable even for a doctor with several years of practice.

Therefore, a computerized Artificial Intelligence (AI) method would allow a remarkable reduction in the expert cardiologist effort and time required, especially useful in the clinical setting for preliminary assessments or even more importantly as a supplementary review tool.

Deep learning methods have recently shown very good results in the field of medical image analysis.
Convolutional neural networks (CNNs) \cite{o2015introduction} are the first deep learning architecture designed specifically for images, they use a process of convolutional layers that automatically learn spatial hierarchies of features. These layers apply filters with small regions of the input data and capture patterns like edges and shapes at different levels of abstraction. CNNs have different layers, convolutional layers for applying filters to the input data, pooling layers  for aggregating features and reducing the size of feature maps, and fully connected layers to use these extracted features for specific tasks. 
While CNN is based on convolutional operations, Transformer \cite{vaswani2017attention} updates the relative significance of some parts of the input through the attention mechanism. Thus, it gives more or less importance to the different words composing a sentence and improves its context and relationship understanding capability. Vision Transformers (ViTs) \cite{dosovitskiy2021imageworth16x16words} are vision-specific and handle image patches as a sequence of tokens like words in a sentence. Such architectures demonstrate high performance in image classification, object detection, and segmentation tasks.

Techniques utilizing CNNs for identifying stenosis from XCA images can be divided into approaches that analyze single frames and those that assess multiple frames. The first method analyze each frame separately \cite{ovalle2022hybrid}\cite{wan2018automated}\cite{cong2019automated}. A very recent prominent study presented a deep learning based methodology for the classification and localization of stenosis in XCA images of 194 patients \cite{cong2019automated}. Their approach used CNNs and Recurrent Neural Networks (RNNs) to achieve good performance on multiple tasks. In particular, the method achieved an Area Under the Curve (AUC) of 0.91 for the classification of both 3-category 
and 2-category stenosis in the right coronary artery (RCA), while in the left coronary artery (LCA), it attained 0.85 for the 3-category and 0.87 for the 2-category classification.  Also, the sensitivity for stenosis detection was 0.72 for the RCA and 0.60 for the LCA, with mean square errors of 69.6 pixels for the RCA and 79.5 pixels for the LCA. It showed impressive performance for both classification and positioning of stenosis, which was the first indication of deep learning models’ capability for boosting the accuracy and reliability of CAD diagnosis. These traditional approaches overlook the temporal dynamics in XCA images, leading to potential difficulties in processing frames with suboptimal image quality

Moreover, Transformers have shown excellent performance in medical imaging tasks because they can capture long-range dependencies and contextual information \cite{JUNGIEWICZ2023120234}. However, their associated computational complexity is a big challenge since it makes them hard to implement and places high requirements on computational resources, hence affecting their practicality, especially in a clinical setup where speed is necessary.

In this study, we develop and evaluate multiple U-shaped segmentation models \cite{ronneberger2015u} to address the challenge of identifying stenosis in a dataset comprising 1500 X-ray angiograms. The majority of these models are based on Mamba \cite{gu2024mambalineartimesequencemodeling} architecture, a novel sequence model that surpasses transformers in terms of linear computational complexity. These models were trained using a dataset \cite{popov2024dataset}, in which experienced doctors identified the stenotic plaques using the SYNTAX Score definition. As per this definition, a stenosis with a thickness greater than 1.5 mm and  50\% or more narrowing is labeled as a coronary lesion.

\section{Related Work}
Stenosis detection from XCA images presents a real challenge given the complex and varied anatomy of coronary arteries, poor image quality, and small size of stenotic areas, which causes class imbalance in the segmentation task.

Recent research has been directed toward developing the segmentation and localization of multivessel coronary artery stenosis from XCA images. Notably, one of the works \cite{bilal2023multivessel} proposed a solution that combined baseline models with ensemble techniques and new post-processing methods. In this paper, an ensemble strategy boosted the results to a position of 5th in the ARCADE challenge, which produced mean F1 scores of 37.69\% in artery segmentation and 39.41\% in stenosis localization. It can deliver intelligent tools for the improvement of CAD diagnosis, guiding interventions, and increasing the accuracy of stent injections in routine clinical practice.
Another leading study \cite{lee2023ssass} proposed an efficient method that integrated data augmentation with a pseudo-label-based semi-supervised learning technique. This approach reached state-of-the-art performance without relying on model ensembles, using a simple model such as YOLOv8 and obtained a 53.6\% F1-score on the ARCADE dataset \cite{popov2024dataset}, which proved the efficiency of the suggested technique in running complex medical image-based tasks for enhancing the accuracy and reliability of CAD diagnosis.
Several studies have also focused on developing advanced architectures and algorithms for the accurate detection of stenosis from XCA images. Among those, this StenUNet \cite{lin2023stenunet} architecture is designed to improve the quality of stenosis detection. This approach obtained an F1-score as high as 0.5348 on a test set only 0.0005 points less than that of the 2nd place, demonstrating its competitiveness in the identification of stenosis accurately.

\section{Data}
The ARCADE Challenge dataset \cite{popov2024dataset}  includes two datasets: one for coronary artery segmentation, and the other for coronary stenosis segmentation. Each dataset consists of 1000 2D X-ray images for training, 200 for validation, and 300 for testing. For this, the first 1200 images were used for training with 5-fold cross-validation, and the rest 300 for testing the final model. All images were standardized at 512×512 pixels, and atherosclerotic plaques were annotated accurately by medical professionals.

\section{Methods}

\subsection{Mamba}
Recently, State Space Models (SSMs) have held great promise \cite{gu2021efficiently}, and the Mamba Models have exploited these developments quite effectively, bringing in important innovations of their own, including the selective scan algorithm. This new model is claimed to be way more efficient and high-performing compared to transformers, which are leading models in recent years. Mamba proposes a new class of models called selective structured state-space models. The key objective of this model family is to retain the transformative power of transformers while achieving linear scaling with sequence length. The paper on Mamba argues that it can retain the power of transformation seen in transformers; however, with linear computational complexity \cite{gu2024mambalineartimesequencemodeling}.

\begin{figure}[htbp]
    \centering
    \fbox{\includegraphics[width=0.47\textwidth]{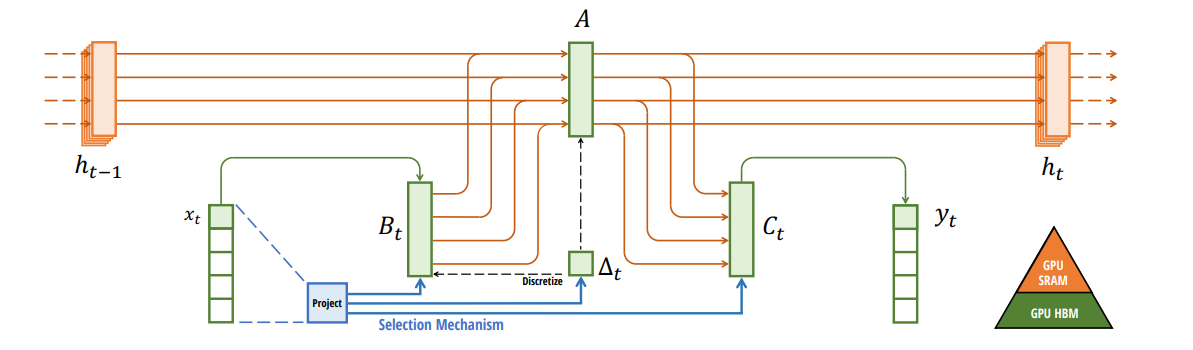}}
    \caption{\small The Mamba architecture\cite{gu2024mambalineartimesequencemodeling}}
    \label{fig1}
\end{figure}

While the methodology that Mamba follows is closest to the recurrent models, what differentiates the approach from others is in terms of the selection mechanism in Figure \ref{fig1}, through which Mamba efficiently selects data based on input dependency, which fixes one major limitation of traditional models. This is what helps Mamba screen off irrelevant information for later use by parametrisation of SSM parameters on the input. Apart from this, Mamba contains hardware-aware algorithms like any other SSM-based model and employs scan operations instead of convolution for computation.

The architecture of Mamba is based on prior SSM architectures, improved by adding the Multilayer Perceptron (MLP) block from transformers, making it a straightforward and efficient model. To sum it up, Mamba has shown itself good in these features:

\begin{enumerate}
    \item {High quality is assured by the selection algorithm used in the model.}
    \item It provides fast training and generation of results since computation and memory scale linearly.
\end{enumerate}

\subsection{Visual State Space Block}
So far, we have seen what Mamba is and how it works. Mamba was created for 1D sequences like text, and in an image, each pixel is not only related to the last or the next pixel. Each pixel is related to other pixels around it to scan most of the information out from our input.

To address this, they introduced selective scan 2D (SS2D) to avoid the direction-sensitive challenge, and they also developed the cross scan module. This does 4 different directions of the scan to the patches \cite{liu2024vmamba}.
The result of each of these cross scans is then fed into Mamba's selective SSMs.
Now that each of our pixels has the information of all directions and surrounding pixels, we will transform all of them to the original 2d form they had before so we can say we made them ready for the next part, which is cross merge. In this step, we will merge them to obtain the final output. For us, this whole process is called SS2D or 2d Selective Scan.

\begin{figure}[h]
    \centering
    \fbox{\includegraphics[width=0.47\textwidth]{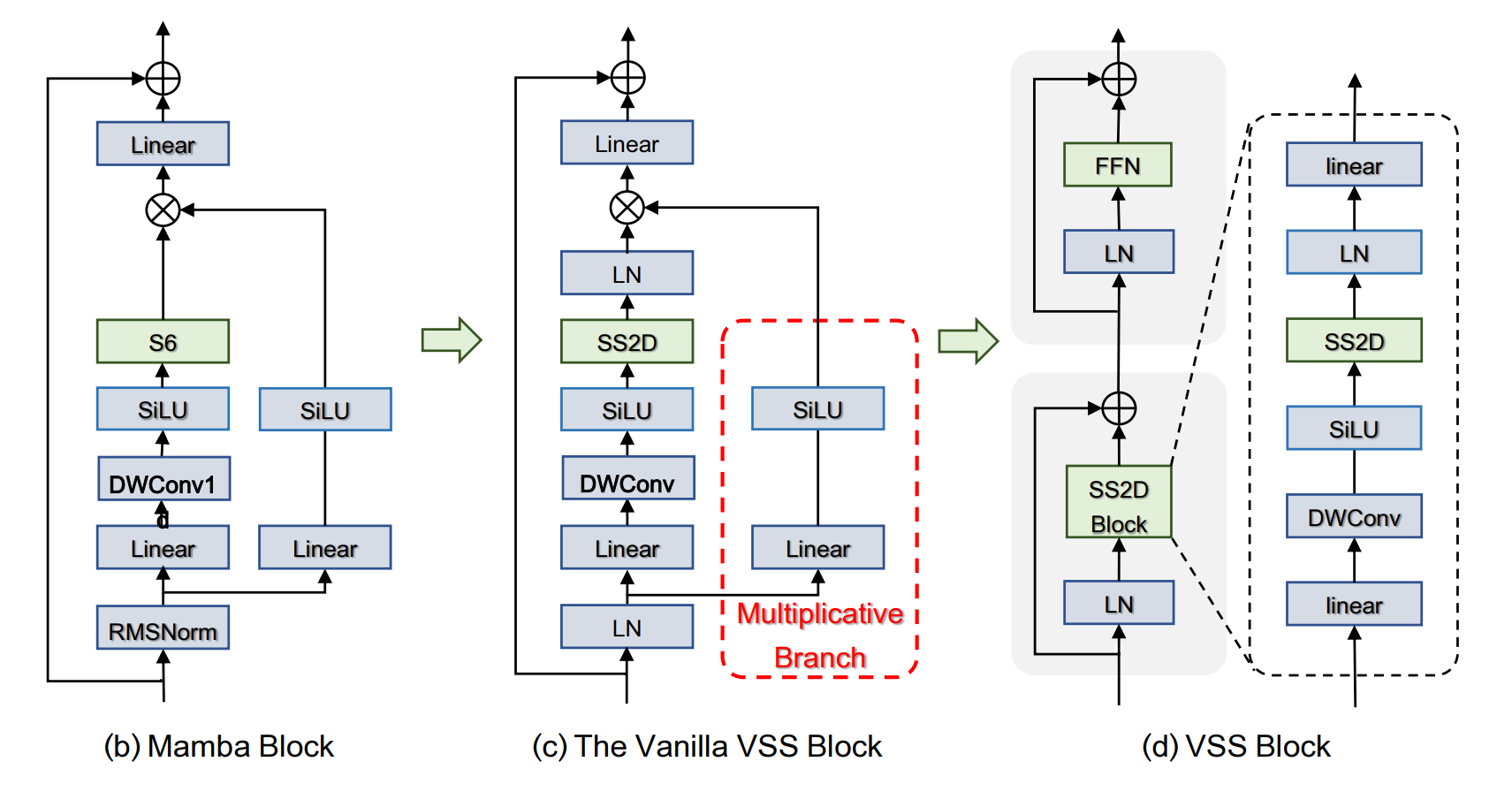}}
    \caption{\small The structure of Mamba and VSS blocks\cite{liu2024vmamba}}
    \label{fig2}
\end{figure}

\begin{figure*}[!t]
    \centering
    \includegraphics[width=\textwidth]{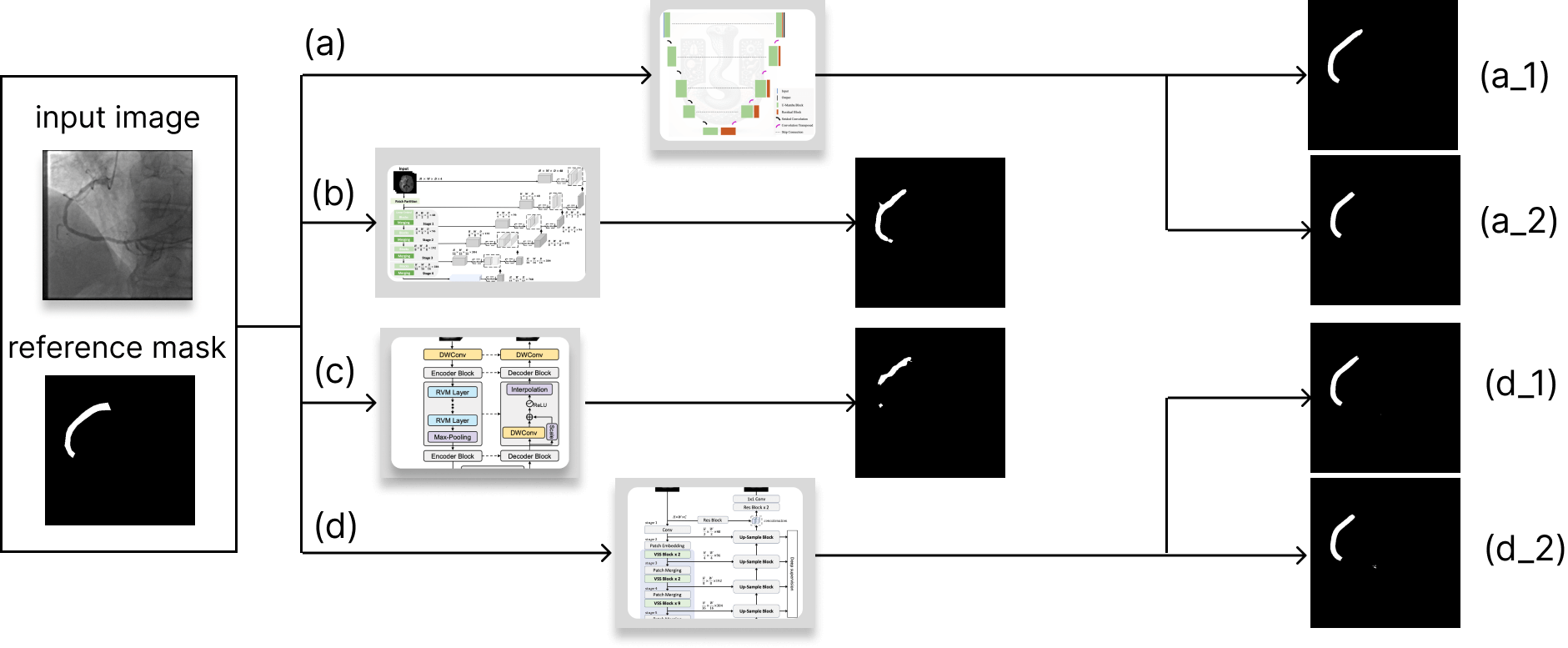}
    \caption{Methodology used in this study. The input images and corresponding masks are fed into various model architectures, and the predictions for each model are displayed in front of the respective model. (a) The U-Mamba architecture \cite{ma2024u}, (b) The Swin UNetR architecture \cite{hatamizadeh2021swin}, (c) The LightM-UNet architecture \cite{liao2024lightm}, (d) The Swin U-Mamba architecture \cite{liu2024swin}. (a\_1) U-Mamba BOT prediction, (a\_2) U-Mamba ENC prediction, (d\_1) Swin U-Mamba prediction, (d\_2) Swin U-Mamba D prediction.}
    \label{fig_combined2}
\end{figure*}

We have two types of Visual State Space (VSS) blocks: Vanilla VSS blocks and VSS blocks. The Vanilla VSS block was introduced by replacing 1D convolution with 2D depth-wise convolution and selective SSMs with SS2D, in addition to a layer of normalization. The new VSS block relies on the SS2D module instead of the S6 module used by traditional Mamba blocks. This simplifies the architecture and provides easier access to global receptive fields and dynamic weights. The new VSS block has fewer components and only has a single network branch with two applied residual modules. Check Figure \ref{fig2} for the architecture of these blocks.\\

\textbf{\textit{U-Mamba:}} 
The first model we will going to discuss is U-Mamba. It has an encoder-decoder network structure, fully capturing all information types within the images. Including Mamba in the architecture makes it efficient and convenient for training.

Each block of the U-Mamba model consists of two successive residual blocks, followed by an SSM-based Mamba block. Much like the above-mentioned VMamba, there are no VSS blocks in the U-Mamba. Instead, it is based on basic S6 Mamba models you can check the architecture in Figure \ref{fig_combined2} \cite{ma2024u}.

There are two variants of U-Mamba: ENC and BOT. U-Mamba makes use of the blocks of U-Mamba in both the encoder and bottleneck, following skip connections to pass information to the decoder. The BOT model incorporates Mamba blocks only in the bottleneck portion of the structure, One of the most striking features of the U-Mamba model is its U-Net-shaped architecture \cite{ronneberger2015u}, which has been highly effective in image segmentation tasks. It is well-known that U-Net and its variants performed very well in image segmentation tasks. It will be quite promising to have a model that could marry Mamba's capabilities with the U-Net encoder-decoder structure.

\textbf{\textit{LightM-UNet:}}
The second model we explored is the lightweight variant of the U-Mamba model, with ×116 parameter and ×21 computational cost drop. Much like the original U-Mamba, this version retains the U-Net structure but is optimized to contain only 5 million parameters, making it very efficient to train on custom datasets.

This is followed by presenting the Residual Vision Mamba Layer (RVM) layer that will help extract deep features of the input. The next fabrication block for this will be carried out for each of the RVM layers before placing a normalization block, followed by the processing block with an SSM-based Mamba model. In addition, some of the modules have depth-wise convolution layers, which help extract the shallow features.The encoder doubles the number of channels in feature maps after every layer and halves the resolution. It is the reverse process of what the decoder does to arrive at the original resolution from the model's output. Notably, the Mamba-based modules are not used in the decoder part of this model. Finally, a softmax at the end yields final segmentation masks from the architecture This means that, in the pursuit of a lightweight model, on the encoder side, it uses blocks based on Mamba only and, in the bottleneck, successive RVM layers, four in number as shown in Figure  \ref{fig_combined2}. These RVM layers retain channel size and resolution and thus provide an efficient and lightweight model \cite{liao2024lightm}.

\textbf{\textit{Swin U-Mamba:}}
The final model we used is Swin-UMamba. This model has an encoder based on Mamba, pre-trained on a large-scale dataset like ImageNet, and could extract features at multiple scales. In this case, the encoder was designed to be for this kind of model, using the V-Mamba Tiny version trained on ImageNet.

Similar to the preceding models, Swin-UMamba is based on a U-shaped structure and follows an encoder-decoder approach with numerous skip connections between the encoder and decoder. VSS blocks are used in the encoder and turn out to be very effective in handling 2D sequences for Mamba with very high accuracy and reduced computational requirements \cite{liu2024swin}.

The Encoder expands the input into four parts, each processed by an S6 basic Mamba block. These result in four different outputs, each giving a different view of the context surrounding a pixel. The decoder is consistent with other models having a U-Net structure and has been successful in segmentation. In addition, the architecture incorporates skip connection features handling so that a further convolution block comes with a residual connection to complete the process. On the other hand, it has an extra segmentation head at the end of each level of decoder to capture deep supervision, as illustrated in Figure \ref{fig_combined2}. They also introduced a variant called Swin-UMamba D, which uses the Mamba blocks in the decoder as well. The number of parameters is fewer, and on segmentation tasks, it is expected to be better.

\textbf{\textit{Swin UNetR:}}
We also used Swin UNetR \cite{hatamizadeh2021swin}, a respected vision transformer based model, for segmentation tasks to assess how transformer models compare to Mamba models in terms of efficiency and power. This comparison is particularly important as our dataset has an imbalance, which typically poses challenges for segmentation models.

Swin UNetR is based on a Swin Unet transformer encoder that can extract features at different resolutions using a shifted window system to handle the self-attention mechanism \cite{cao2022swin}. Swin Transformers, also known as Shifted Window Transformers is an architecture intended to improve the efficiency and effectiveness of ViTs in computer vision assignments. Unlike ViTs that process information from the entire image simultaneously Swin Transformers divide the input image into non overlapping local windows each processed independently \cite{liu2021swin}. This window oriented approach reduces complexity while maintaining performance. To capture context and enable interaction between windows Swin Transformers employ a shifting mechanism where windows are shifted in layers for cross window communication. This hierarchical design allows Swin Transformers to effectively model long range dependencies and multi scale features making them particularly beneficial for tasks, like image classification, object detection and semantic segmentation.

Swin UNetR utilizes a design based on U-Net incorporating an encoder and decoder in a U-shaped network structure. The connection, between the encoder and decoder is facilitated by skip connections that pass data seamlessly. Residual blocks are integrated into both the encoder and decoder to preserve information, from stages, For an overview of the structure please consult Figure \ref{fig_combined2}.

\begin{figure}[htbp]
    \centering
    \fbox{\includegraphics[width=0.47\textwidth]{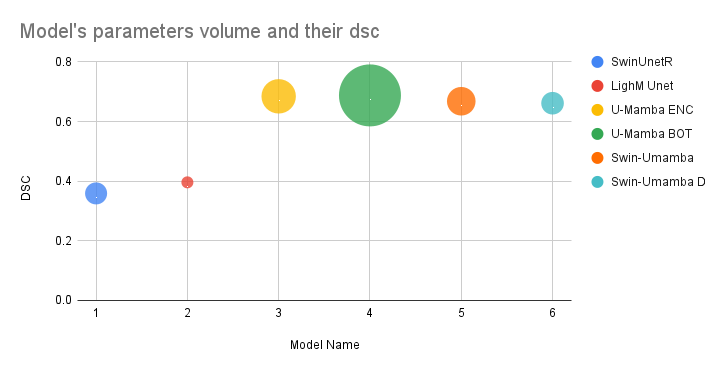}}
    \caption{\small Visualization of each model's performance on the ARCADE dataset \cite{popov2024dataset}. The Y-axis of this chart shows their performance you have to check the central point of each circle for it and circle diameters represent the number of parameters used in each of these models.}
    \label{fig3}
\end{figure}

\begin{figure}[ht]
    \centering
    \begin{minipage}{0.47\textwidth}
        \includegraphics[width=\textwidth]{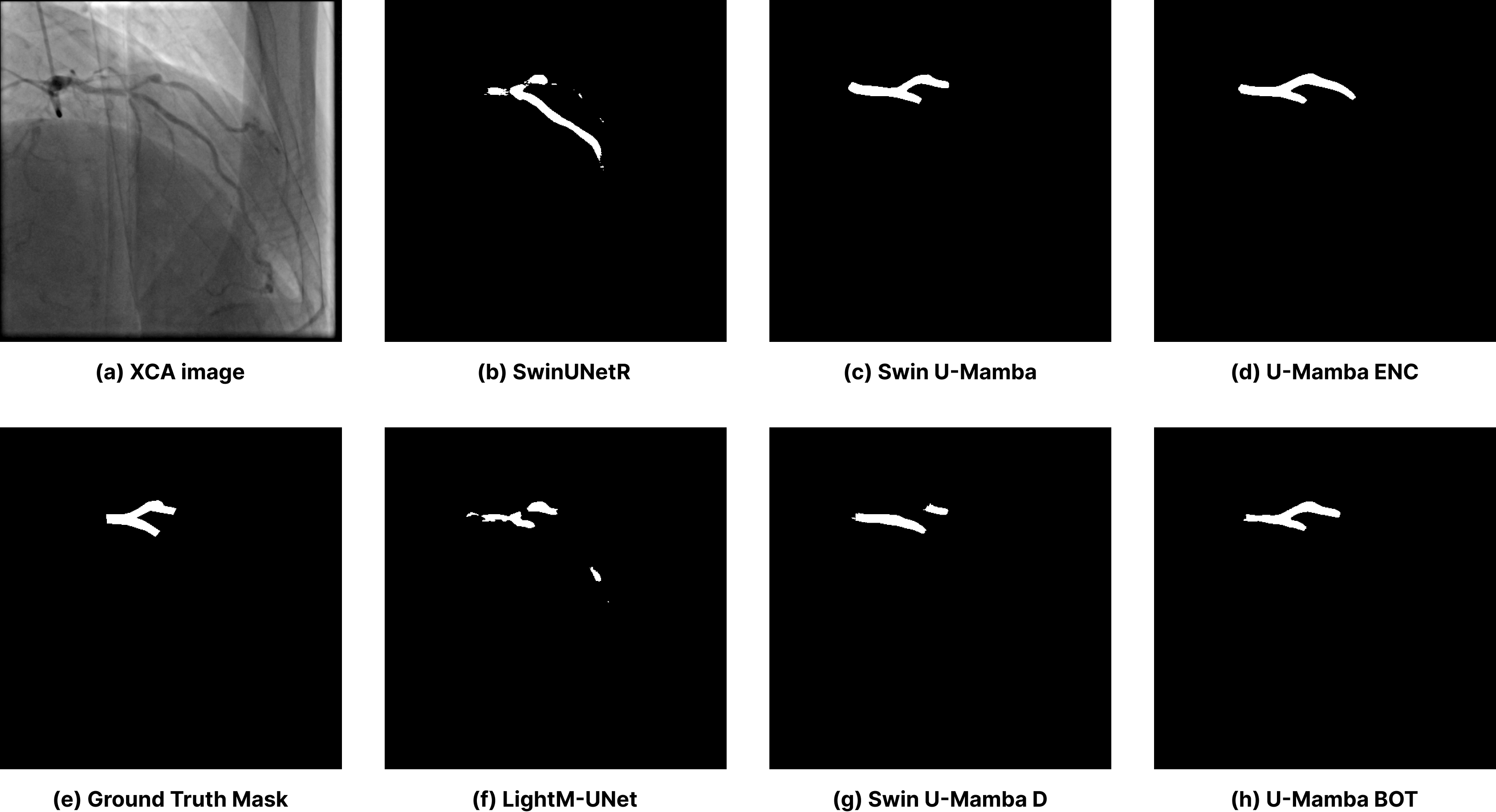}\vspace{1mm}
        \includegraphics[width=\textwidth]{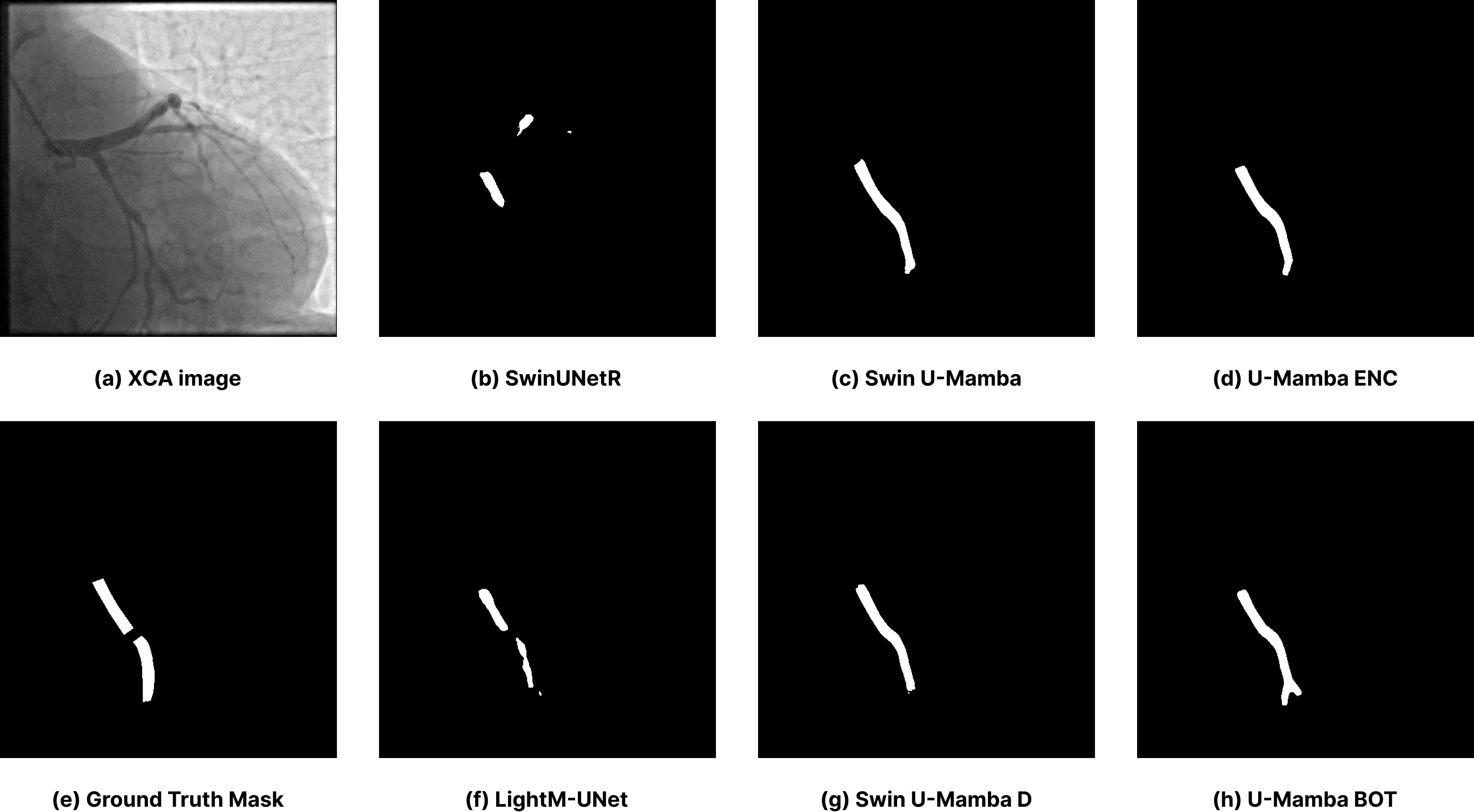}\vspace{1mm}
        \includegraphics[width=\textwidth]{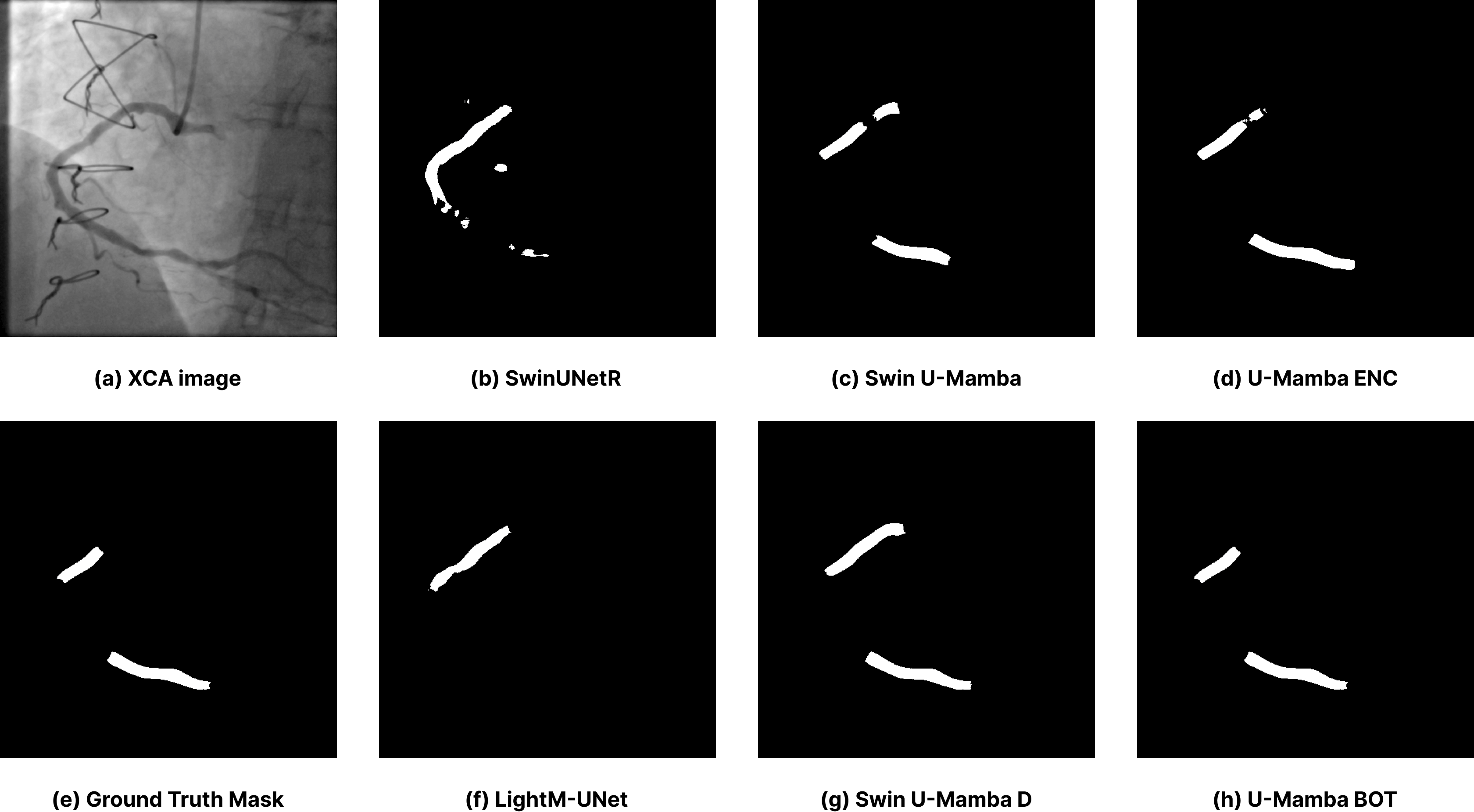}\vspace{1mm}
        \includegraphics[width=\textwidth]{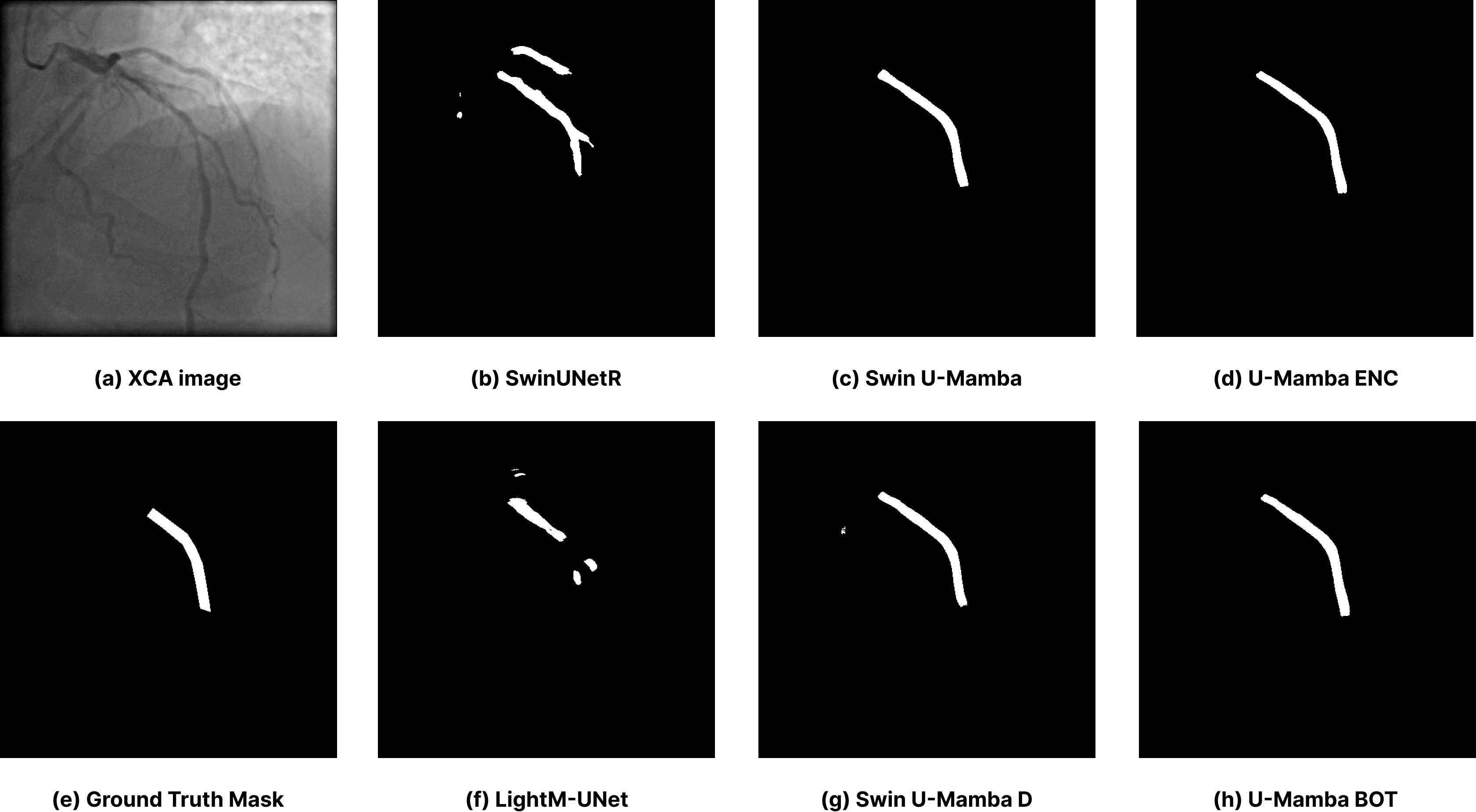}
    \end{minipage}%
    \caption{Prediction of different models on test dataset.}
    \label{fig:overall_caption}
\end{figure}

\section{Results}
In this section, we will evaluate the models against certain metrics. The most important metrics for us here were precision and recall so let's dive into it.

Precision and recall are indeed very basic metrics to be used while evaluating a segmentation model, more so in the case of an imbalanced dataset. Precision is, in fact, the positive predictive value; that is the ratio between instances we truly predicted as positive and all of the cases were predicted positively by our model.

Recall, also referred to as sensitivity or the true positive rate, answers the question: "How many of the true positive cases did the model exactly identify?"

Probably one of the main drawbacks to using accuracy is that it is not very informative. Sometimes, it can be hard to balance precision and recall because improving one usually involves deteriorating the other. As such, we also report an F1 score, which is the harmonic mean of precision and recall to balance between these two metrics. The F1 score is appropriate for segmentation tasks; many other benchmarks also went for it on this task. The F1 metric of our models is reported in Table \ref{tab:model_comparison}.

We tested some of the models for this task and presented their results in Table  \ref{tab:model_comparison}. In this figure, we can see that while U-Mamba BOT had the best performance, LightM-UNet was the one with the fewest parameters. Another point to be highlighted here is the Swin U-Mamba D, which has the Mamba blocks on both sides. It broke some of the benchmark results, showing that having Mamba blocks on either side of the model makes it effective and efficient for tasks, thus supporting its claim in their original paper\cite{bilal2023multivessel}\cite{lee2023ssass}\cite{lin2023stenunet}.

As we outlined, you can refer to the metrics we described to see the trade-off between performance and efficiency of similar tasks. Figure  \ref{fig3} is added here to give a general comparison among all the models concerning the mentioned metrics: precision, recall, and F1 score. This should be able to help one make an appropriate choice and even help in picking up the most appropriate model according to the need.

\begin{table}[ht]
\centering
\caption{Model Performance Comparison}
\label{tab:model_comparison}
\begin{tabular}{lcccc}
\toprule
\textbf{Model Name} & \textbf{Precision} & \textbf{Recall} & \textbf{F1-Score} & \textbf{Params} \\
\midrule
Ensemble Learning \cite{bilal2023multivessel} & -- & -- & 0.3941 & -- \\
Sten UNet \cite{lin2023stenunet} & -- & -- & 0.5348 & -- \\
SSASS \cite{lee2023ssass} & -- & -- & 0.5699 & -- \\
Swin UNetR & 0.4912 & 0.2829 & 0.359 & 25M \\
LightM-UNet & 0.4893 & 0.3326 & 0.396 & 5M \\
Swin-Umamba D & 0.6869 & 0.6378 & 0.6614 & 27M \\
Swin-Umamba & 0.6887 & 0.6488 & 0.6682 & 60M \\
U-Mamba ENC & 0.7113 & 0.6618 & 0.6857 & 104M \\
U-Mamba BOT & 0.6992 & 0.6769 & 0.6879 & 500M \\
\bottomrule
\end{tabular}
\end{table}

\section{Conclusion}

We trained five different variants of the Mamba-based model and one variant of the Swin Transformer-based model with a major use of the U-Net architecture in the detection of stenosis location in CAD. In this regard, our U-Mamba BOT model achieved the best F1 score and recall, while the highest precision was obtained by using the U-Mamba ENC model. This might mean that the U-Mamba ENC model probably forecasts fewer true positive stenosis pixels but also fewer false positives, which can be an advantage for this kind of task.

These models, based on Mamba, have shown very good performance for this challenging task on this imbalanced dataset and hence can be considered reliable for further vision tasks.






%

\bibliographystyle{plain}  
\bibliography{main}  

\end{document}